# Analysis of Critical State Response in Thin Films by ac Susceptibility Measurements

A. Youssef, Z. Švindrych, J. Hadač, and Z. Janů

*Abstract*—Low frequency ac magnetic response of the Nb thin film in perpendicular oscillating applied fields is detected by continuously reading SQUID magnetometer. Harmonic analysis of the temperature dependence of the nonlinear ac susceptibility gives excellent agreement with the susceptibility calculated on basis of model of the hysteretic critical state in 2D disk. This complete analytical model relates to Bean's critical state described by the critical depinning current density. We map the experimental and model data and we trace the temperature dependence of the critical current density. The model fits up to temperature $0.9997T_c$ above which nonlinear response turns to linear one. This technique is also suitable for characterization of HTS thin films.

## I. INTRODUCTION

THERE are only few complete analytical models of magnetic response of type II superconductors in critical state to external magnetic field. These models are limited to Bean's critical state model and restricted to geometry of infinitely long cylinder or slab in parallel field or infinite stripes in perpendicular field. Recently, in 1993, model for 2D disk in perpendicular field was introduced. In this model the control parameter is ratio of ac field amplitude and product of film thickness and critical depinning current density. This model does not involve any temperature dependence. On the other hand, measurement of temperature dependence of ac susceptibility is common experimental technique. Then, mapping of the model data to experimental data gives temperature dependence of the critical current density. Numerical methods are needed for other geometry of the sample. This applies also for non quasi-static critical state when vortices diffuse or flow.

Because of high aspect ratio of a film in a perpendicular applied magnetic field vortices may nucleate at edges and enter film interior even in weak fields, smaller than is Earth's magnetic field (about 50 µT).

The case of 2D film with Bean critical state in perpendicular external field concerns many applications like HTS YBCO and $MgB_2$ power cables but also thin Nb films for superconducting electronics, because the Nb is one of only three elementary type-II superconductors.

## II. EXPERIMENTAL

The Nb film of thickness of 250 nm was deposited by dc magnetron sputtering in Ar gas on 400 nm thick silicon-dioxide buffer layer which was grown by a thermal oxidation of silicon single crystal wafer [1]. The film is poly-crystalline with texture of a preferred orientation in the (110) direction and is highly tensile. Grain size is about 100 nm. The square samples of 5 mm × 5 mm in dimensions were cut out from the 3-inch wafer.

Magnetic moment of the sample was detected by continuously reading SQUID magnetometer, which is similar to one described in [2]. In this magnetometer, unlike commercial ones, the sample is placed stationary in one of gradiometer coils. The SQUID output signal is proportional to the difference of magnetic fluxes in the gradiometer coils caused by the sample's magnetic moment. Applied homogeneous field is created by a superconducting solenoid which operates in non-persistent mode. The solenoid is supplied from voltage driven current source. The SQUID output and voltage monitoring the supply current to the solenoid as well as the voltages driving the current source are converted by over-sampling delta-sigma modulating converters. Delta-sigma converters are inherently linear, low total harmonic distortion, and high signal-to-noise ratio. These features allow acquire or generate signals with high accuracy and fidelity and consequently perform high resolution spectral (harmonic) analysis.

The sample was mounted by Apiezon grease on a flat surface of the cylindrical sapphire holder. GaAlAs diode temperature sensor was mounted on the opposite side. The holder was suspended on both thermally and electrically insulating support which was connected to stainless steel tube. This insert was placed in anti-cryostat filled with $^4$He gas at atmospheric pressure. Temperature of the anti-cryostat was controlled by Si diode temperature sensor, Lake Shore temperature controller, and resistive wire heater. High permeability cylinder and superconducting lead can shield external field. Residual field may be compensated.

Measurements were done in time-varying continually oscillating field, $H(t)=H_{ac}\sin(\omega_0 t)$, while the sample was

Manuscript received August 28, 2007. This work was supported by the Czech Science Foundation under Grant No. 102/05/0942.

A. Youssef and Z. Švindrych are with the Charles University in Prague, Faculty of Mathematics and Physics, Ke Karlovu 3, 121 16 Prague 2, Czech Republic.

J. Hadač is with the Czech Technical University in Prague, Faculty of Nuclear Sciences and Physical Engineering, Břehová 7, 115 19 Prague 1, Czech Republic

Z. Janů is with the Institute of Physics of Czech Academy of Sciences, v.v.i., Na Slovance 2, 182 21 Prague 8, Czech Republic (corresponding author e-mail: janu@fzu.cz).

cooled or warmed slowly. Applied frequencies, $\omega_0/2\pi$, were in the range from 1.5625 to 12.5 Hz. The magnetization signal, $M(t)$, and signal monitoring the applied field, $H(t)$, were Fourier transformed in real time (via discrete fast Fourier transform) to complex frequency spectra, $M(\omega)$ and $H(\omega)$. The fundamental frequency and harmonics of the complex susceptibility were obtained from the magnetization deconvolved by field

$$\chi_n = \frac{H^*(n\omega_0) \cdot M(n\omega_0)}{H^*(n\omega_0) \cdot H(n\omega_0)} \tag{1}$$

to eliminate shift of analyzed data segment. The asterisks mark the complex conjugated numbers. The coefficient of $n^{th}$ harmonic of the field spectrum is

$$H(n\omega_0) = |H(\omega_0)|(\arg H(\omega_0))^n . \tag{2}$$

### III. CRITICAL STATE MODEL IN 2D FILM

The Bean's model of the critical state, on which the model is constructed, posses essential simplifications: a) The critical state is isotropic, $|\mathbf{j}| = j_c$ in flux penetrated regions, where $|\mathbf{B}| \neq 0$, while $|\mathbf{j}| = 0$ in non-penetrated regions. b) $j_c$ is independent from the local flux density $\mathbf{B}$. c) The surface does not constitute barrier. d) The critical state is quasistatic.

Mikheenko and Kuzovlev constructed the model in 2D disk-shaped superconductor by analogy to that of a long cylinder and found the complete analytical solutions of field and current patterns in the thin film disk-shaped type-II superconductors in perpendicular time-varying periodic applied magnetic fields [3].

This model was corrected and extended by Zhu et al. who showed that since the current density cannot change discontinuously in interior of the film, magnitude of the current density decreases continuously in vortex-free annulus from $j_c$ to zero in center of the disk [4]. Because of large aspect ratio of 2D disk in perpendicular field, parallel with $z$ axis, the flux density bends around the disk and shielding currents flow over the entire surface of the disk. The magnitude of $\partial B_r/\partial z$ is much larger than magnitude of $\partial B_z/\partial r$ for weak external fields except at the center of the disk; the shielding current mainly comes from the term $\partial B_r/\partial z$, in contrast to the case of a long cylindrical sample. The model is constrained for the film thickness $d \leq 2\lambda$, where $\lambda$ is the London flux penetration length, to assure that the circulating currents in the film plane may be treated as having uniform density in the thickness direction. Also, the external field is assumed to be weak enough so that the critical current density in the film is independent of the local density of trapped vortices, i.e. $j_c d$ = const. Zhu et al. gave the complete analytical solution to current density $\mathbf{j}(r)$ and magnetization $\mathbf{M}(r)$ profiles in 2D thin film disk, to the magnetic moment, to both initial (virgin) magnetization and for time-varying field, and to effective magnetic susceptibility.

Further work on was carried out by Clem and Sanchez who showed that the model may be extended for either $d \geq \lambda$ or, if $d < \lambda$, that $\Lambda = 2\lambda^2/d \ll R$, where $\Lambda$ is the two-dimensional screening length [5]. They calculated, in addition, analytical solution to complex ac-susceptibility (fundamental frequency and harmonics) components and gave its approximative behavior. They concluded that for finite applied fields the annular region where the current density is $j_c$ never fills the entire disk, and the critical-state flux-density profiles never penetrate all the way to the center, where $B_z$ remains equal to zero. In real thin films, however, the above 2D approach breaks down and $B_z$ becomes nonzero when the vortex-free radius $a$ approaches the largest of the quantities $d$, $\lambda$, and $\Lambda$. They summarized that independently of geometry of the sample the hysteretic critical-state behavior is that during quasistatic changes of an applied magnetic field, vortices move and thus the local flux density $\mathbf{B}$ changes wherever the magnitude of the current density $\mathbf{j}$ (assumed perpendicular to the vortices) exceeds the critical value $j_c$. On the other hand, in long samples the local current density $\mathbf{j}$ can change only where the flux density changes while in a film changes in the applied field induce screening currents $\mathbf{j}$ to flow not just at the edges but throughout the film. However, these currents do not cause flux motion or a change in the perpendicular component of the magnetic flux density unless the magnitude of $\mathbf{j}$ exceeds $j_c$.

For temperature approaching $T_c$ both $H_{c1}$ and $j_c$ turn to zero. Then even in weak applied fields may be $H \gg H_{c1}$ and one may put $\mathbf{B} = \mu_0\mathbf{H}$. In this case the quasistatic Bean's model fails and dynamic model should be used instead. The flux changes are considered to be slow or quasistatic if the magnitude of the electric field $\mathbf{E}$ induced by the moving magnetic flux is small by comparison with $\rho_{FF}j_c$, where $\rho_{FF}$ is the flux-flow resistivity. Under these conditions, the magnitude of the induced current density $\mathbf{j}$ is always very close to the critical current density $j_c$ in regions where $\mathbf{B}$ is changing. If flux changes are not quasistatic the pinning and flux creep may be modeled by a nonlinear resistivity law $\mathbf{E} = \rho(j)\mathbf{j}$ [6], [7]. If pinning is strong or if the magnetization is reversible, the magnetic response is nonlinear. The linear response may be observed at finite temperatures where thermally assisted depinning of vortices occurs. Application of such models, however, requires numerical methods to solve differential equations [8]. For given geometry and model $\mathbf{E}(\mathbf{j})$ the current density may be calculated by time integration of a nonlocal diffusion equation.

### IV. RESULTS AND DISCUSSION

#### A. Meeting of Model Assumptions

Primarily we assess meeting of the model assumptions: a) The film thickness, $d = 250$ nm, is larger than is the flux penetration length for Nb, $\lambda(0) = 40$ nm, thus the first Clem and Sanchez condition, $d > \lambda$, is fulfilled. However, the $\lambda$ may become larger than the thickness near $T_c$. Considering Gorter-Casimir or Ginzburg-Landau model for temperature dependence of $\lambda$, the $\lambda(T) \approx d$ for $T \approx 0.99$-$0.999\ T_c$. However, in this case, the second Clem and Sanchez assumption on the two-dimensional screening length, $\Lambda$, is fulfilled as $\Lambda^2 \ll dR/2 \approx 250$ nm $\times$ 2.5 mm $\times$ 0.5, i.e. $\Lambda \ll 18$ μm. b) The model is for disk geometry but our samples are of the square geometry. Here we refer to Brandt's calculations which show that the difference between both geometries is only 1% [9]. c) The ac susceptibilities were measured in low frequency field from 1.5625 to 12.5 Hz. As the susceptibilities do not change with frequency, field may be considered quasi-static. d) Appropriateness of the model may be reviewed briefly looking on the peak value of the measured $\chi''$, because the model gives value $\chi'' \approx 0.241$, which occurs for $H_{ac}/H_d = 1.942$, where $H_d = j_c d/2$ is the characteristic field. This value is independent on any particular temperature dependence of the critical current density.

### B. Calculation of Magnetization Curves

The magnetization curves were computed using the complete analytical solution to the magnetization of the 2D disk-shaped superconductor in time-varying applied field [4], [5]. When the external field varies periodically with amplitude $H_{ac}$, the magnetization is

$$M_\mu = \mu \chi_0 H_{ac} S\left(\frac{H_{ac}}{H_d}\right) \pm \chi_0 (H_{ac} \mu H) S\left(\frac{H_{ac} \mu H}{2 H_d}\right), \quad (3)$$

where $M_-$ and $M_+$ is for decreasing (upper signs) and increasing applied field, respectively. The function $S(x)$ is defined as

$$S(x) = \frac{1}{2x}\left[\arccos\left(\frac{1}{\cosh x}\right) + \frac{\sinh |x|}{\cosh^2 x}\right]. \quad (4)$$

Nonlinear complex ac susceptibilities were calculated from the hysteresis loops of the magnetization using coefficients of Fourier transformed $M(\varphi)$ and $H(\varphi)$ arrays in the same way the experimental susceptibility was calculated.

### C. Mapping of experimental data to model data

To match the experimental and model susceptibilities, we plot the model $\chi$ versus negative reciprocal value of $2H_{ac}/j_c d$ scaled by parameter $c$, i.e. $-c(j_c d/2H_{ac})$, unlike commonly used plot versus $2H_{ac}/j_c d$ [5], [10], and experimental $\chi$ in a convenient temperature scale, $(T-T_c)/T_c$. The scaling, $j_c(T) \sim ((T_c-T)/T_c)^\alpha$, where $\alpha > 0$, meets expected behavior with $j_c \to 0$ as $T \to T_c$. The fundamental frequency experimental ac susceptibility for frequency 1.5625 Hz and model ac susceptibility are shown in Fig. 1. The experimental susceptibilities were measured in sequence: warming in field $\mu_0 H_{ac} = 10$ µT; cooling in 5 µT; cooling in 2 µT; warming in 1 µT. The temperature rate was 0.1 K/min.

We have found that the linear dependence, $j_c(T)=j_{c0}(T_c-T)/T_c$, where $j_{c0} = 2H_{ac}/cd$, with only two fitting parameters, $c$ and $T_c$, gives excellent agreement for $\mu_0 H_{ac} = 10$ and 5 µT. For $\mu_0 H_{ac} < 5$ µT, the peak in $\chi''$ is smaller than 0.241 and disappears with decreasing amplitude of the ac field.

The values of $T_c$, $c$, temperature $T_p$ at which a maximum $\chi_{1p}''$ of $\chi_1''$ occurs, and calculated $j_{c0}$, for set of amplitudes $H_{ac}$ are listed in Table 1. We note that $j_{c0}$ is assumed to change with temperature but not with $H_{ac}$ otherwise the Bean's model would be inapplicable. As the field is increased, the $T_c$ lowers, in agreement with expected suppression of the superconductivity by applied field. Concurrently, the transition region broadens and the absorption peak shifts to lower temperature. The dependence $T_c(H_{ac})$ is approximately linear and $T_c \to 9.186$ K for zero field.

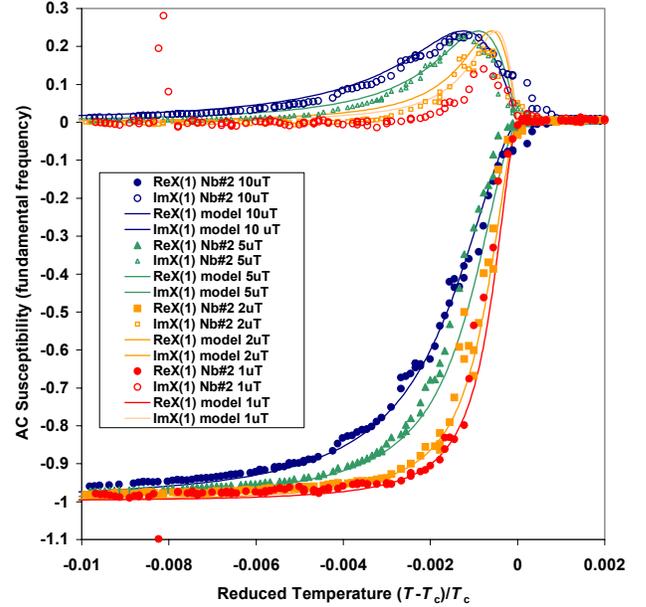

Fig. 1. Real $\chi_1'$ and imaginary $\chi_1''$ parts of the fundamental frequency complex susceptibility. Marks show the measured temperature dependence for $\mu_0 H_{ac} = 10, 5, 2,$ and 1 µT plotted as $[(T-T_c)/T_c, \chi(T)]$. Solid curves show the model data plotted as $[c(H_d/H_{ac}), \chi(H_{ac}/H_d)]$. Both dependencies are mapped by fitting $c$ and $T_c$ only.

TABLE 1 DEPENDENCE OF CHARACTERISTIC PARAMETERS

| $\mu_0 H_{ac}$ (µT) | $T_c$ (K) | $c$ | $T_p$ (K) | $\chi_{1p}''$ | $J_{c0}$ (MA/cm$^2$) |
|---|---|---|---|---|---|
| 10 | 9.165 | 0.0025 | 9.155 | 0.239 | 2.6 |
| 5 | 9.175 | 0.0018 | 9.164 | 0.224 | 1.8 |
| 2 | 9.180 | 0.0012 | 9.173 | 0.205 | 1.1 |
| 1 | 9.185 | 0.001 | 9.178 | 0.141 | 0.64 |

The experimental and model ac susceptibility plotted in log x-axis scale for $\mu_0 H_{ac} = 10$ µT are shown in Fig. 2. The model evidently fits the experimental data up to temperature $3 \times 10^{-4}$ $T_c$ below the critical temperature. This interval represents ~ 3 mK on common temperature scale.

For $\mu_0 H_{ac} = 0.5$ µT, by increasing frequency from 1.5625, 3.125, 6.25, and 12.5 Hz both $T_p$ and $T_c$ change only within temperature resolution. In normal state, the absorption increases linearly with the applied field frequency $\omega/2\pi$. The nonlinear response of the superconductor when pinning is present is reflected in appearance of higher harmonics in the susceptibility. The odd harmonics, 3$^{rd}$ and 5$^{th}$, of both experimental and model susceptibilities are shown in Fig. 3. As the symmetry of magnetization hysteresis curves is odd, $M(H)=-M(-H)$, the even harmonics of ac susceptibility must be zero. The 2$^{nd}$ and 4$^{th}$ harmonics of the experimental susceptibility are shown in Fig. 4. For easy confrontation, the vertical scales were left intentionally the same as in Fig. 3. The 2$^{nd}$ harmonic is smaller than 3$^{rd}$ and 4$^{th}$ is smaller than 5$^{th}$.

Nonzero values of even harmonics near below $T_c$ are consequence of experimental conditions – continuous warming or cooling.

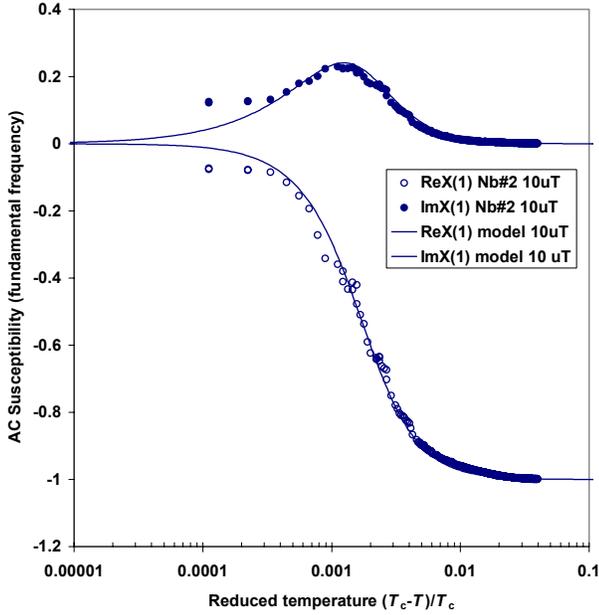

Fig. 2. Real $\chi_1'$ and imaginary $\chi_1''$ parts of the fundamental frequency complex ac susceptibility for $\mu_0 H_{ac}$ = 10 μT. The marks show measured dependence and the curves show model dependence.

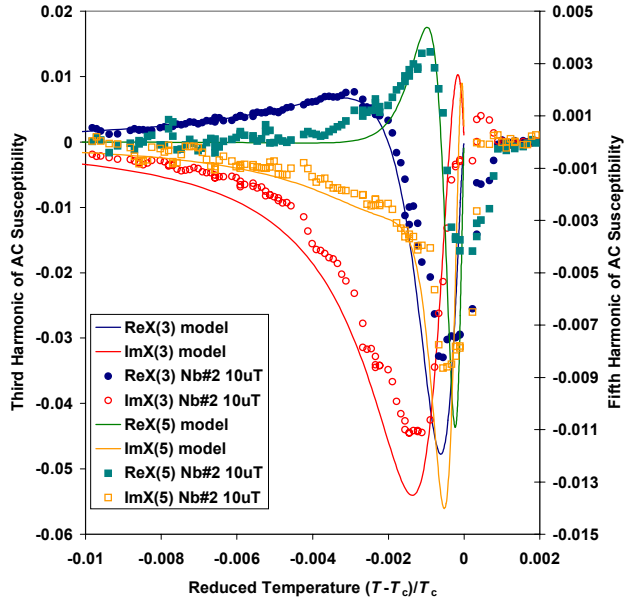

Fig. 3. Real and imaginary parts of odd (3$^{rd}$ and 5$^{th}$) harmonics of complex ac susceptibility. The marks show measured temperature dependence for $\mu_0 H_{ac}$ = 10 μT. The curves show model dependence.

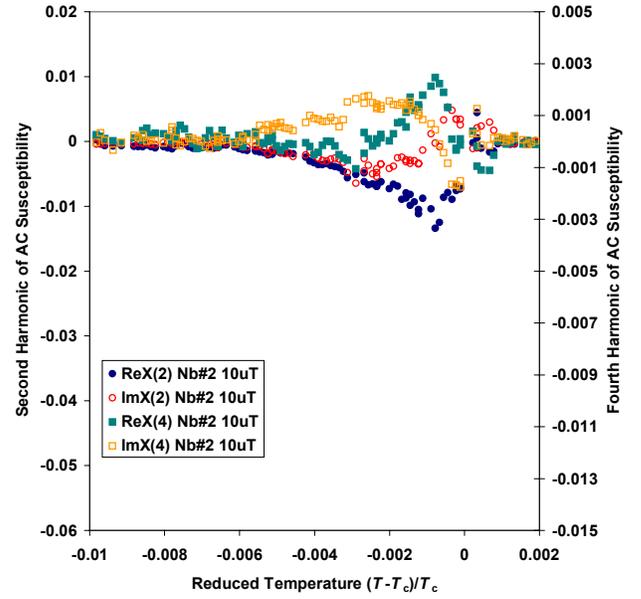

Fig. 4. Real and imaginary parts of even (2$^{nd}$ and 4$^{th}$) harmonics of complex ac susceptibility. The marks show measured temperature dependence for $\mu_0 H_{ac}$ = 10 μT. All even harmonics of the model susceptibility are zero because of symmetry of the magnetization hysteresis curves.

## V. Conclusion

In summary, we have measured temperature dependence of the magnetization of Nb thin film in slowly varying perpendicular applied field. Both fundamental-frequency and harmonics of the experimental ac susceptibility agree well with ac susceptibility computed on basis of the complete analytical model with the Bean's critical state in 2D superconducting disk.

## Acknowledgment

The authors are grateful to T. May, M. Grajcar, F. Soukup, and R. Tichý for stimulating discussion and help with experiment.